
\magnification \magstep1
\raggedbottom
\openup 4\jot
\voffset6truemm
\headline={\ifnum\pageno=1\hfill\else
\hfill {\it Boundary terms for massless fermionic fields}
\hfill \fi}
\rightline {August 1993, SISSA Ref. 137/93/A, DSF preprint 93/27}
\centerline {\bf BOUNDARY TERMS FOR MASSLESS}
\centerline {\bf FERMIONIC FIELDS}
\vskip 1cm
\centerline {\bf Giampiero Esposito$^{1,2}$,
Hugo Morales-T\'ecotl$^{3}$ and
Giuseppe Pollifrone$^{1}$}
\vskip 1cm
\centerline {\it ${ }^{1}$Istituto Nazionale di Fisica Nucleare}
\centerline {\it Mostra d'Oltremare Padiglione 20, 80125 Napoli, Italy;}
\centerline {\it ${ }^{2}$Dipartimento di Scienze Fisiche}
\centerline {\it Mostra d'Oltremare Padiglione 19, 80125 Napoli, Italy;}
\centerline {\it ${ }^{3}$Scuola Internazionale Superiore di Studi Avanzati}
\centerline {\it Via Beirut 2-4, 34013 Trieste, Italy.}
\vskip 1cm
\noindent
{\bf Abstract.} Local supersymmetry leads to boundary conditions
for fermionic fields in one-loop quantum cosmology involving
the Euclidean normal $_{e}n_{A}^{\; \; A'}$ to the boundary and a pair
of independent spinor fields $\psi^{A}$ and
${\widetilde \psi}^{A'}$. This paper studies the corresponding
classical properties, i.e. the classical boundary-value problem
and boundary terms in the variational problem. If
$\sqrt{2} \; {_{e}n_{A}^{\; \; A'}} \; \psi^{A}
\mp {\widetilde \psi}^{A'} \equiv \Phi^{A'}$ is set to zero
on a 3-sphere bounding flat Euclidean 4-space, the modes of the
massless spin-${1\over 2}$ field multiplying harmonics having
positive eigenvalues for the intrinsic 3-dimensional Dirac operator
on $S^{3}$ should vanish on $S^{3}$. Remarkably, this coincides with
the property of the classical boundary-value problem when spectral
boundary conditions are imposed on $S^3$ in the massless case.
Moreover, the boundary term in the action functional is proportional
to the integral on the boundary of $\Phi^{A'} \; {_{e}n_{AA'}}
\; \psi^{A}$.
\vskip 100cm
\noindent
Locally supersymmetric boundary conditions have been recently
studied in quantum cosmology to understand its one-loop
properties. They involve the normal to the boundary and the field
for spin ${1\over 2}$, the normal to the boundary and the
spin-${3\over 2}$ potential for gravitinos, Dirichlet conditions
for real scalar fields, magnetic or electric field for
electromagnetism, mixed boundary conditions for the 4-metric
of the gravitational field (and in particular Dirichlet conditions
on the perturbed 3-metric). The aim of this letter is to describe
the corresponding classical properties in the case of massless
spin-${1\over 2}$ fields.

For this purpose, we consider flat Euclidean 4-space bounded
by a 3-sphere of radius $a$. The spin-${1\over 2}$ field,
represented by a pair of independent spinor fields $\psi^{A}$
and ${\widetilde \psi}^{A'}$, is expanded on a family of
3-spheres centred on the origin as [1-3]
$$
\psi^{A}={\tau^{-{3\over 2}}\over 2\pi}
\sum_{n=0}^{\infty}\sum_{p=1}^{(n+1)(n+2)}
\sum_{q=1}^{(n+1)(n+2)} \alpha_{n}^{pq}
\Bigr[m_{np}(\tau)\rho^{nqA}+{\widetilde r}_{np}(\tau)
{\overline \sigma}^{nqA}\Bigr]
\eqno (1)
$$
$$
{\widetilde \psi}^{A'}={\tau^{-{3\over 2}}\over 2\pi}
\sum_{n=0}^{\infty}\sum_{p=1}^{(n+1)(n+2)}
\sum_{q=1}^{(n+1)(n+2)} \alpha_{n}^{pq}
\Bigr[{\widetilde m}_{np}(\tau){\overline \rho}^{nqA'}
+r_{np}(\tau)\sigma^{nqA'}\Bigr]
\; \; \; \; .
\eqno (2)
$$
With our notation, $\tau$ is the Euclidean-time coordinate,
the $\alpha_{n}^{pq}$ are block-diagonal matrices with
blocks $\pmatrix {1&1\cr 1&-1\cr}$, the $\rho-$ and
$\sigma$-harmonics obey the identities described in [1,3].
Last but not least, the modes $m_{np}$ and $r_{np}$ are
regular at $\tau=0$, whereas the modes ${\widetilde m}_{np}$
and ${\widetilde r}_{np}$ are singular at $\tau=0$ if the
spin-${1\over 2}$ field is massless. Bearing in mind that the
harmonics $\rho^{nqA}$ and $\sigma^{nqA'}$ have positive
eigenvalues ${1\over 2}\Bigr(n+{3\over 2}\Bigr)$ for the
3-dimensional Dirac operator on the bounding $S^3$ [3], the
decomposition (1-2) can be re-expressed as
$$
\psi^{A}=\psi_{(+)}^{A}+\psi_{(-)}^{A}
\eqno (3)
$$
$$
{\widetilde \psi}^{A'}={\widetilde \psi}_{(+)}^{A'}
+{\widetilde \psi}_{(-)}^{A'}
\; \; \; \; .
\eqno (4)
$$
In (3-4), the $(+)$ parts correspond to the modes $m_{np}$
and $r_{np}$, whereas the $(-)$ parts correspond to the singular
modes ${\widetilde m}_{np}$ and ${\widetilde r}_{np}$,
which multiply harmonics having negative eigenvalues
$-{1\over 2}\Bigr(n+{3\over 2}\Bigr)$ for the 3-dimensional
Dirac operator on $S^3$. If one wants to find a classical
solution of the Weyl equation which is regular
$\forall \tau \in [0,a]$, one is thus forced to set to zero
the modes ${\widetilde m}_{np}$ and ${\widetilde r}_{np}$
$\forall \tau \in [0,a]$ [1]. This is why, if one requires the
local boundary conditions [3]
$$
\sqrt{2} \; {_{e}n_{A}^{\; \; A'}} \; \psi^{A}
\mp {\widetilde \psi}^{A'}=\Phi^{A'}
\; {\rm on} \; S^{3}
\eqno (5)
$$
such a condition can be expressed as [3]
$$
\sqrt{2} \; {_{e}n_{A}^{\; \; A'}} \; \psi_{(+)}^{A}
=\Phi_{1}^{A'}
\; {\rm on} \; S^{3}
\eqno (6)
$$
$$
\mp {\widetilde \psi}_{(+)}^{A'}=\Phi_{2}^{A'}
\; {\rm on} \; S^{3}
\eqno (7)
$$
where $\Phi_{1}^{A'}$ and $\Phi_{2}^{A'}$ are the parts of the
spinor field $\Phi^{A'}$ related to the ${\overline \rho}$-
and $\sigma$-harmonics respectively. In particular, if
$\Phi_{1}^{A'}=\Phi_{2}^{A'}=0$
on $S^3$ as in [2,3], one finds
$$
\sum_{n=0}^{\infty}\sum_{p=1}^{(n+1)(n+2)}
\sum_{q=1}^{(n+1)(n+2)}\alpha_{n}^{pq}
\; m_{np}(a) \; {_{e}n_{A}^{\; \; A'}}
\; \rho_{nq}^{A}=0
\eqno (8)
$$
$$
\sum_{n=0}^{\infty}\sum_{p=1}^{(n+1)(n+2)}
\sum_{q=1}^{(n+1)(n+2)}\alpha_{n}^{pq}
\; r_{np}(a)
\; \sigma_{nq}^{A'}=0
\eqno (9)
$$
where $a$ is the 3-sphere radius. Since the harmonics
appearing in (8-9) are linearly independent, these
relations lead to $m_{np}(a)=r_{np}(a)=0$ $\forall n,p$.
Remarkably, this simple calculation shows that the classical
boundary-value problems for regular solutions of the Weyl
equation subject to local or spectral conditions on $S^3$
share the same property provided $\Phi^{A'}$ is set to zero
in (5): the regular modes $m_{np}$ and $r_{np}$ should vanish
on the bounding $S^3$.

To study the corresponding variational problem for a massless
fermionic field, we should now bear in mind that the
spin-${1\over 2}$ action
functional in a Riemannian 4-geometry takes the form [2,3]
$$
I_{E}={i\over 2}\int_{M}\Bigr[{\widetilde \psi}^{A'}
\Bigr(\nabla_{AA'}\psi^{A}\Bigr)
-\Bigr(\nabla_{AA'}{\widetilde \psi}^{A'}\Bigr)
\psi^{A}\Bigr]\sqrt{{\rm det} \; g} \; d^{4}x+{\widehat I}_{B}
\; \; \; \; .
\eqno (10)
$$
This action is {\it real}, and the factor $i$ occurs
by virtue of the convention for Infeld-van der Waerden symbols
used in [2,3].
In (10) ${\widehat I}_{B}$ is a suitable boundary term, to be
added to ensure that $I_{E}$ is stationary under the boundary
conditions chosen at the various components of the boundary
(e.g. initial and final surfaces, as in [1]). Of course, the
variation $\delta I_{E}$ of $I_{E}$ is linear in the
variations $\delta \psi^{A}$ and $\delta {\widetilde \psi}^{A'}$.
Defining $\kappa
\equiv {2\over i}$ and $\kappa {\widehat I}_{B} \equiv I_{B}$,
variational rules for anticommuting spinor fields lead to
$$ \eqalignno{
\kappa \Bigr(\delta I_{E}\Bigr)&=
\int_{M}\Bigr[2 \delta {\widetilde \psi}^{A'}
\Bigr(\nabla_{AA'}\psi^{A}\Bigr)\Bigr]
\sqrt{{\rm det} \; g} \; d^{4}x
-\int_{M}\Bigr[\Bigr(\nabla_{AA'}{\widetilde \psi}^{A'}
\Bigr)2 \delta \psi^{A}\Bigr]
\sqrt{{\rm det} \; g} \; d^{4}x \cr
&-\int_{\partial M}\Bigr[{_{e}n_{AA'}}
\Bigr(\delta {\widetilde \psi}^{A'}\Bigr)
\psi^{A}\Bigr]\sqrt{{\rm det} \; h} \; d^{3}x
+\int_{\partial M}\Bigr[{_{e}n_{AA'}}
{\widetilde \psi}^{A'}\Bigr(\delta \psi^{A}\Bigr)
\Bigr] \sqrt{{\rm det} \; h} \; d^{3}x \cr
&+\delta I_{B}
&(11)\cr}
$$
where $I_{B}$ should be chosen in such a way that its
variation $\delta I_{B}$ combines with the sum of the two terms
on the second line of (11) so as to specify what is fixed
on the boundary (see below). Indeed, setting
$\epsilon = \pm 1$ and using the boundary conditions (5)
one finds
$$
{_{e}n_{AA'}}{\widetilde \psi}^{A'}
={\epsilon \over \sqrt{2}}\psi_{A}
-\epsilon \; {_{e}n_{AA'}}\Phi^{A'}
\; {\rm on} \; S^{3}
\; \; \; \; .
\eqno (12)
$$
Thus, anticommutation rules for spinor fields [1] show that the
second line of equation (11) reads
$$ \eqalignno{
\delta I_{\partial M} & \equiv
-\int_{\partial M}\Bigr[\Bigr(\delta {\widetilde \psi}^{A'}
\Bigr){_{e}n_{AA'}}\psi^{A}\Bigr]
\sqrt{{\rm det} \; h} \; d^{3}x
+\int_{\partial M}\Bigr[{_{e}n_{AA'}}{\widetilde \psi}^{A'}
\Bigr(\delta \psi^{A}\Bigr)\Bigr]
\sqrt{{\rm det} \; h} \; d^{3}x \cr
&=\epsilon \int_{\partial M}
{_{e}n_{AA'}}\Bigr[\Bigr(\delta \Phi^{A'}\Bigr)
\psi^{A}-\Phi^{A'}\Bigr(\delta \psi^{A}\Bigr)
\Bigr] \sqrt{{\rm det} \; h} \; d^{3}x
\; \; \; \; .
&(13)\cr}
$$
Now it is clear that setting
$$
I_{B} \equiv \epsilon \, \int_{\partial M}
\; \Phi^{A'} {_{e}n_{AA'}}
\; \psi^A \sqrt{ {\rm det} \; h} \; d^3x
\; \; \; \; ,
\eqno (14)
$$
enables one to specify $\Phi^{A'}$ on the boundary, since
$$
\delta \Bigr[ I_{\partial M} + I_{B} \Bigr] =
2 \epsilon \int_{\partial M} {_{e}n_{AA'}}
\Bigr( \delta \Phi^{A'} \Bigr) \psi^A \sqrt{{\rm det }\; h}\; d^3x
\; \; \; \; .
\eqno (15)
$$
Hence the action integral (10) appropriate for our boundary-value
problem is
$$ \eqalignno{
I_{E}&={i\over 2}\int_{M}\Bigr[{\widetilde \psi}^{A'}
\Bigr(\nabla_{AA'}\psi^{A}\Bigr)
-\Bigr(\nabla_{AA'}{\widetilde \psi}^{A'}\Bigr)
\psi^{A}\Bigr]\sqrt{{\rm det} \; g} \; d^{4}x \cr
&+{i\epsilon \over 2} \, \int_{\partial M}
\; \Phi^{A'} {_{e}n_{AA'}}
\; \psi^A \sqrt{ {\rm det} \; h} \; d^3x
\; \; \; \; .
&(16)\cr}
$$
Note that, by virtue of (5), equation (13) may also be
cast in the form
$$
\delta I_{\partial M}
={1\over \sqrt{2}}\int_{\partial M}
\Bigr[{\widetilde \psi}^{A'}\Bigr(\delta \Phi_{A'}\Bigr)
-\Bigr(\delta {\widetilde \psi}^{A'}\Bigr)\Phi_{A'}
\Bigr] \sqrt{{\rm det} \; h} \; d^{3}x
\; \; \; \; ,
\eqno (17)
$$
which implies that an equivalent form of $I_{B}$ is
$$
I_{B} \equiv {1\over \sqrt{2}}
\int_{\partial M}{\widetilde \psi}^{A'}
\; \Phi_{A'} \sqrt{{\rm det} \; h}
\; d^{3}x
\; \; \; \; .
\eqno (18)
$$

The local boundary conditions studied at the classical level
in this paper, have been applied to one-loop quantum cosmology
in [2-4]. Interestingly, our work seems to add evidence in favour
of quantum amplitudes having to respect the properties of the
classical boundary-value problem. In other words, if fermionic
fields are massless, their one-loop properties in the presence
of boundaries coincide in the case of spectral [1,3,5] or local
boundary conditions [2-4], while we find that classical modes
for a regular solution of the Weyl equation obey the same
conditions on a 3-sphere boundary with spectral or local
boundary conditions, provided the spinor field $\Phi^{A'}$
of (5) is set to zero on $S^{3}$. We also hope that the
analysis presented in Eqs. (10)-(18) may clarify the
spin-${1\over 2}$
variational problem in the case of local boundary conditions
on a 3-sphere (cf. the analysis in [6] for pure gravity).
\vskip 1cm
\leftline {\bf Acknowledgments}
\vskip 1cm
\noindent
We are indebted to George Ellis for encouraging our work.
The first author has learned the two-component spinor
analysis of fermionic fields from Peter D'Eath.
The second author thanks for financial support the
Ministero Italiano per l'Universit\`a e la Ricerca
Scientifica e Tecnologica, and CONACyT
(Reg. No. 55751).
Financial support by the Istituto Nazionale di Fisica
Nucleare is gratefully acknowledged from the third
author.
\vskip 10cm
\leftline {\bf REFERENCES}
\vskip 1cm
\item {1.}
P. D. D'Eath and J. J. Halliwell, {\it Phys. Rev. D}
{\bf 35}, 1100 (1987).
\item {2.}
P. D. D'Eath and G. Esposito, {\it Phys. Rev. D}
{\bf 43}, 3234 (1991).
\item {3.}
G. Esposito, {\it Quantum Gravity, Quantum Cosmology
and Lorentzian Geometries}, Lecture Notes in Physics,
New Series m: Monographs, Vol. m12 (Springer-Verlag,
Berlin, 1992).
\item {4.}
A. Y. Kamenshchik and I. V. Mishakov, {\it Phys. Rev. D}
{\bf 47}, 1380 (1993).
\item {5.}
P. D. D'Eath and G. Esposito, {\it Phys. Rev. D}
{\bf 44}, 1713 (1991).
\item {6.}
J. W. York, {\it Found. Phys.} {\bf 16}, 249 (1986).
\bye